\documentclass[12pt]{article}
\usepackage{amssymb}

\newcommand{\p}{\partial}
\newcommand{\pdot}{\partial \!\cdot \!}
\usepackage[utf8]{inputenc}
\usepackage{hyperref}
\usepackage{amsmath,,calrsfs}
\usepackage{amsmath,amssymb}
\usepackage{graphicx}
\usepackage{calrsfs}
\usepackage{breqn}
\DeclareMathAlphabet{\pazocal}{OMS}{zplm}{m}{n}

\usepackage{tikz}
\usepackage{ragged2e}
\usepackage{indentfirst}
\usepackage[nottoc,numbib]{tocbibind}
\usepackage{booktabs, multirow} 
\usepackage{soul}
\usepackage[brazil,english]{babel}

\newcommand{\be}{\begin{equation}}

\def \del{\partial}
\newcommand{\ee}{\end{equation}}
\newcommand{\lbt}{\overline{\Lambda}^T}
\newcommand{\lbdd}{\overline{\Lambda}^{\ddagger}}
\newcommand{\lb}{\overline{\Lambda}}

\numberwithin{equation}{section}

\newcommand{\nn}{\nonumber}

\newcommand{\no}{\noindent}

\def\bea{\begin{eqnarray}}
\def\eea{\end{eqnarray}}
\def\beq{\begin{eqnarray}}
\def\eeq{\end{eqnarray}}

\def \eps {\epsilon }
\def \lap {\nabla^2}
\let\mathcal\pazocal
\begin{document}

\title{\textbf{A partially broken Fronsdal model for massless higher-spin particles of integer spin}}

	\author{D. Dalmazi\footnote{denis.dalmazi@unesp.br} \ and  B. dos S. Martins\footnote{bruno.s.martins@unesp.br}\\
		\textit{{UNESP - Campus de Guaratinguetá - DFI }}} 
	\date{\today}
	\maketitle

\begin{abstract}

By introducing arbitrary parameters in the usual Fronsdal model, we find a region in the parameters space away from the ``Fronsdal point'' where we still have an irreducible description of massless particles of arbitrary integer spin $s\ge 3$. The higher spin gauge symmetry is further constrained by a vanishing double divergence condition on the traceless gauge parameter: $\p\cdot\p\cdot\bar\Lambda=0$. Remarkably, it does not introduce extra propagating gauge invariants. We demonstrate that we only have spin-$s$ helicity states as propagating modes for arbitrary integer $s\ge 3$. For the simplest $s=3$ case  we present a gauge invariant proof while for $s\ge 4$ we use a light-cone gauge. The reduction in the gauge symmetry allows for more general source couplings when compared to the Fronsdal model.


\end{abstract}
\newpage

\section{Introduction}

The issue of higher spin (HS) particles has a long history \cite{dirac,fp}, see the review works \cite{snow,rakibur_r,sagnoti_r1,vasiliev_r,sagnoti_r} and the book \cite{book}. Here we are mainly concerned with massless particles of integer spin-$s$.  In a metric-like formulation they can be described by a rank-s totally symmetric tensor. As the spin grows, the number of field components rapidly increases. Since in $D=4$ dimensions we must end up  with only two  helicities ($\pm s$),  we need an enlarging number of gauge symmetries as well in order to get rid of redundant field components. 

The basic model for a Lorentz covariant  irreducible description of a free integer spin-$s$ massless particle is the Fronsdal model \cite{fronsdal}. The gauge symmetry in \cite{fronsdal} is given by a traceless HS kind of  diffeomorphism\footnote{We work in flat Minkowski spacetime with $\eta = \text{diag}(-,+,+,\dots,+)$ and $D\ge 3$ dimensions. Symmetrizations have no weights, e.g., $T_{(\mu\nu)} = T_{\mu\nu} + T_{\nu\mu}$. We use in part of the work the compact notation of \cite{fs1}, where symmetrizations are implicit: $\partial^k h = \partial_{(\mu_1}\cdots \partial_{\mu_k} h_{\mu_{k+1} \dots \mu_{k+s})}\,$, $\,\eta^k h = \eta_{(\alpha_1\beta_1}\cdots \eta_{\alpha_k\beta_k} h_{\mu_{1} \dots \mu_{s})}$. The dotted terms $\partial \cdot$ and $\eta \cdot$ represent contractions: $(\partial \cdot)^k h = \partial^{\mu_1} \cdots \partial^{\mu_k} h_{\mu_1 \dots \mu_k\mu_{k+1}\cdots\mu_s}$ and $(\eta \cdot)^k h = \eta^{\mu_1 \mu_2} \cdots \eta^{\mu_{2k-1}\mu_{2k}} h_{\mu_1\cdots\mu_{2k-1}\mu_{2k}\mu_{2k+1} \cdots \mu_s} \equiv h^{[k]}$. Primes denote lower $k$ traces, e.g., $h^{'}=\eta\cdot h$ and $h^{[2]} = {h}''$. A single bar indicates a traceless field ($\eta^{\mu_1\mu_2} \bar{h}_{\mu_1\mu_2 \dots \mu_s} = 0$).} $\delta_F h = \p\, \overline{\Lambda}$, where $\overline{\Lambda}_{\mu_1\cdots\mu_{s-1}}$ is a traceless symmetric tensor of rank-(s-1). There is another model \cite{sv} where the traceless gauge parameter is further restricted to be spacetime transverse: $\p\cdot \lbt =0$. However, in this case, a traceless Weyl-like symmetry is required, such that the total gauge transformation is $\delta_{SV} h = \p \lbt + \eta\, \overline{\sigma} $. At the end both $\delta_F h$ and $\delta_{SV} h$ contain the same number of independent gauge parameters. 

Beyond free particles, the subject of consistent interactions of HS particles has proved to be rather challenging. It is constrained by several ``No-Go'' theorems, see, e.g., \cite{porrati,bbs}. It is expected that the less gauge symmetries one requires, the easier one builds up consistent interacting vertices. So it is important to ask oneself whether one could still have an irreducible description of massless integer spin particles with less gauge symmetry. 

The aim of the present work is to show that we can further reduce the gauge symmetry while still having only $\pm s$ helicities as propagating modes for arbitrary integer spin $s\ge 3$ in $D=4$. This is carried out by modifying the Fronsdal model and the gauge symmetry by imposing a vanishing double divergence condition ($\p\cdot \p \cdot \lb=0$) on the traceless gauge parameter. The corresponding theories we name here partially broken Fronsdal (PBF) models. It is also possible to modify the model of \cite{sv} reducing the symmetry by demanding a transverse traceless Weyl parameter ($\p\cdot\overline{\sigma}^T=0$). 

Here we use a method for the identification of the particle's spectrum of a gauge theory based on certain gauge invariants known as Bardeen variables \cite{Jaccard} which can be constructively built up out of the gauge transformations as in \cite{da-hs}. In the next section we introduce the method via the Fronsdal models of $s=2$ (linearized Einstein-Hilbert) and $s=3$. In section 3 we show how the method saves us  from several steps in the spectrum identification in some cases which include reducible models of  spin-2 and spin-3. The latter one naturally leads us to the spin-3 PBF model. We present a gauge invariant proof that such model has only spin-$3$ helicities as propagating modes. We define PBF models for arbitrary integer $s\ge 3$ in section 4. We look at its equations of motion and argue that the model describes the expected number of propagating modes (two in $D=4$). In subsection 4.3 we choose the light cone gauge and prove that only $\pm s$ helicities propagate in $D=4$. In section 5 we draw our conclusions and comment on future works.

\section{\bf Spin-2 and spin-3 models via gauge invariants}

As an introduction to the higher spin cases ($s \ge 3$) we first apply  the gauge invariant method of spectrum identification  in the spin-2 case. At $s=2$ the Fronsdal model coincides with the linearized Einstein-Hilbert (LEH) theory. It is invariant under linearized diffeomorphisms,
\be \delta \, h_{\mu\nu} = \p_{\mu}\, \eps_{\nu} + \p_{\nu}\, \eps_{\mu}  . \label{ddiff} \ee 

\no Splitting space and time indices we can use $D$
 out of the $D(D+1)/2$ equations (\ref{ddiff}) to obtain the gauge parameters as functions of the gauge transformations: $\eps_{\mu}(\delta\, h_{\alpha\beta})$. Plugging back the obtained gauge parameters in 
(\ref{ddiff}) we obtain $N_I\equiv D(D+1)/2-D$ invariance equations: $\delta \, I^{Diff}_A =0$, where $A=1,2, \cdots, N_I$. The quantities $I^{Diff}_A$ represent diffeomorphisms (Diff henceforth) invariants. Explicitly, see \cite{nstd} for a recent derivation, $I_A^{Diff} = \{I^{tt}_{ij},I_{i}^t,I_{00},I\}$, where\footnote{The expression $(1/\nabla^2)$  represents the corresponding Green's function, for instance, in $D=4$ we have $(1/\lap)f (\vec{x},t)= \int\, d^3r^{\prime} f(\vec{x}^{\prime},t)/(4\pi\vert r-r^{\prime}\vert)$.  The single index ``$t$'' stands for space  transverse ($\p_{j}v_{j}^t=0$) quantities while ``$tt$'' for space transverse and traceless ones ($\p_{i}v_{ij}^{tt}=0=\delta^{ij}v_{ij}^{tt}$), $i,j=1,2,\cdots,D-1$. We use the index ``$T$'' for space-time transverse fields ($\p^{\mu}v_{\mu}^T$=0).},

\be	I^{tt}_{ij}= \left\lbrack\theta_{ik}\theta_{jl}-\frac{\theta_{ij}\theta_{kl}}{(D-2)}\right\rbrack h_{kl}  \quad ;\qquad  I_{i}^t=\theta_{ij}\left[h_{0j}-\frac{\partial_{k}\dot{h}_{jk}}{\nabla^2}\right]\ \ . \label{diffinv}\ee

\be  I_{00} =h_{00}+\frac{\omega_{ij}\ddot{h}_{ij}}{\nabla^2}-2\frac{\partial_j \dot{h}_{0j}}{\nabla^2}\quad ; \quad I = \theta_{ij}\, h_{ij} \quad , \label{diffinvs} \ee 

\no with the space transverse projection operator given by $\theta_{ij}= \delta_{ij} - \omega_{ij}$ where $\omega_{ij}= \p_i\p_j/\lap$.

In terms of $I^{Diff}_A$, also known as Bardeen variables \cite{Jaccard}, the LEH theory becomes
\bea \mathcal{L}_{LEH}&=& \frac{1}{2} h^{\alpha\beta}\, \Box \,  h_{\alpha\beta}+(\p^{\mu} h_{\mu\beta})^2+  h\, \p^{\mu}\p^{\nu} h_{\mu\nu}-\frac{1}{2} h\,  \Box h \label{leh}\\
&=&I_{ij}^{tt}\frac{\Box}{2}I_{ij}^{tt}-I_{j}^{t}\,\nabla^2\, 
I_{j}^{t}-\frac{(D-3)}{2(D-2)}I\, \Box \, I + I\, \nabla^2 \, I_{00}. \label{lehinv} \eea

\no By decomposing the fundamental field potential $h_{\mu\nu}$ in terms of $SO(D-1)$ tensors, also called an helicity decomposition, it is possible to show \cite{nstd} that each invariant $I_A^{Diff}$ can be considered as an independent field coordinate. So we have one equation  of motion for each invariant. Therefore, on shell from (\ref{lehinv}) we have  $D(D-3)/2$ massless modes: $\Box\, I_{ij}^{tt}=0 $. In $D=4$ they correspond to the propagating $\pm 2$ helicities. Moreover,  throughout this work we assume vanishing fields at infinity such that the Laplace equation $\nabla^2 f=0$ has only the trivial solution, $f=0$. Consequently the other equations of motion from (\ref{lehinv}) lead, in the absence of sources, to vanishing invariants: $I_j^t=0=I=I_{00}$. 

A natural spin-3 analogue of the diffeomorphisms (\ref{ddiff}) would be
\be \label{3diff} \delta\,  h_{\mu\nu\alpha} = \partial_{(\mu}\Lambda_{\nu\alpha)}, \ee

\no where  $ \Lambda_{\nu\alpha} = \Lambda_{\alpha\nu} $. We can follow the same pattern of the spin-2 case and deduce spin-3 Diff invariants directly from (\ref{3diff}).  It is convenient to decompose the gauge parameters $\Lambda_{\mu\nu}$ in $SO(D-1)$ tensors, 
  \bea \nn \Lambda_{00} &=& A \quad ; \quad \Lambda_{0j} = B_j^t + \partial_j C\ ; \\
       \label{quasefim} \Lambda_{ij} &=& D_{ij}^{tt} + \partial_{(i}D_{j)}^t + \nabla^2\theta_{ij}d + \partial_i \partial_j \tilde{d} . \label{Lij}
      \eea

      \no Plugging back in the $D(D+1)(D+2)/6$ equations (\ref{3diff}) we can use $D(D+1)/2$ of them in order to obtain $\Lambda_{\mu\nu}=\Lambda_{\mu\nu}( \delta\,  h_{\rho\alpha\beta}) $. Decomposing also the field potential $h_{\mu\nu\alpha}$ in $SO(D-1)$ tensors,
\bea \nn h_{000} &=& \rho \quad ;\quad h_{00j} = \gamma_j^t + \partial_j \Gamma\ ; \\
       h_{0ij} &=& \psi_{ij}^{tt} + \partial_{(i}\psi_{j)}^t + \nabla^2 \theta_{ij}\psi + \partial_i \partial_j \tilde{\psi}\ ;  \label{helicscampos3} \\
     \nn h_{ijk} &=& \xi^{tt}_{ijk} + \partial_{(i}\xi_{jk)}^{tt} + \theta_{(ij}\tilde{\xi}^t_{k)} + \partial_{(i}\partial_j\xi^t_{k)} + \nabla^2\partial_{(i}\theta_{jk)}\xi + \partial_i\partial_j\partial_k\tilde{\xi}, \eea
     
\no we can write $\Lambda_{\mu\nu}( \delta\,  h_{\rho\alpha\beta})$ as
     \bea \nn A &=& \delta(\Gamma - \dot{\tilde{\psi}} + \frac{1}{3}\ddot{\tilde{\xi}}) \quad ; \quad B_j^t = \delta(\psi_j^t - \frac{1}{2}\dot{\xi_j^t})\ ; \\ \label{ce} C &=& \frac{1}{2}\delta(\tilde{\psi} - \frac{1}{3}\dot{\tilde{\xi}}) \quad ; \quad D_{ij}^{tt} = \delta (\xi^{tt}_{jk}) \quad ; \quad 
     D_j^t = \frac{1}{2}\delta(\xi^t_k)\ ; \\
     \nn d &=& \delta(\xi) \quad ; \quad \tilde{d} =\frac{1}{3}\delta(\tilde{\xi}).
     \eea     

    \no From (\ref{quasefim}) and (\ref{ce}) back in (\ref{3diff}) we obtain $N_I^{Diff}=D(D+1)(D+2)/6 - D(D+1)/2$ invariance equations $\delta \, I_A^{Diff}=0 $ where the spin-3 Diff invariants are given by:
    \bea  I_{ijk}^{tt} &=& \xi_{ijk}^{tt} \quad ; \quad I_{ij}^{tt} =\dot{\xi_{ij}^{tt}} - \psi_{ij}^{tt}\ ; \label{invss3} \\
     I_j^t &=& \gamma_j^t - 2\dot{\psi_j}^t + \ddot{\xi_j}^t \quad ; \quad \tilde{I_j}^t = D \tilde{\xi}_j^t\ ; \label{ijota}\\
     I &=& \rho - 3\dot{\Gamma} + 3\ddot{\tilde{\psi}} - \dddot{\tilde{\xi}} \quad ; \quad \tilde{I} = \dot{\xi} - \psi.
    \label{itil}\eea

  \no  It is important to mention that both decompositions (\ref{quasefim}) and (\ref{helicscampos3}) are invertible as far as we consider vanishing fields at infinity. So we can always write $I_A^{Diff}=I_A^{Diff}(h_{\mu\nu\alpha})$ for all invariants, in particular, 
  \be I_{ijk}^{tt} = \theta_{il}\theta_{jm}\theta_{kn} \, h_{lmn} - \frac{\theta_{(ij}\theta_{k)m}(\theta\cdot h)_m}{D}, \label{iijk} \ee

  \no where $(\theta\cdot h)_m=\theta_{ij}\, h_{ijm} $.

It turns out that there is no local Lorentz invariant free spin-3 theory symmetric under the unconstrained Diffs (\ref{3diff}). However, if we only require a traceless parameter (trDiff henceforth),
 \be \label{dtrd} \delta\,  h_{\mu\nu\alpha} = \partial_{(\mu}\overline{\Lambda}_{\nu\alpha)} \quad ; \quad \eta^{\mu\nu}\overline{\Lambda}_{\mu\nu}=0 \quad  \ee

 \no and a Lorentz covariant second order in derivatives field theory, we arrive at the spin-3 Fronsdal \cite{fronsdal} model up to trivial local field redefinitions,
\be  \pazocal{L}_F = -\frac{1}{2}(\partial_\rho h_{\mu\nu\alpha})^2 + \frac{3}{2}(\partial^\mu h_{\mu\nu\alpha})^2 + 3\partial^\mu \partial^\nu h_{\mu\nu\alpha} h^\alpha + \frac{3}{2}(\partial_\rho h_\nu)^2 + \frac{3}{4}(\partial^\mu h_\mu)^2 , \label{f3} \ee
\no where $h_{\alpha}=\eta^{\mu\nu}h_{\mu\nu\alpha}$. It becomes clear from the previous examples (\ref{ddiff}) and (\ref{3diff}), that the number of independent gauge invariants equals the number of fields which transform under gauge transformations minus the number of independent gauge parameters. Generically, 
\be N_I = N(h) - N(\Lambda) . \label{ni} \ee

\no So the traceless condition  reduces by one unit the number of independent gauge parameters, leading to one extra gauge invariant. From $\eta^{\mu\nu} \bar{\Lambda}_{\mu\nu} = -A + \nabla^2[\tilde{d} + (D-2)d] = 0$ in (\ref{ce})  we obtain $\delta\, I_{tr}=0$, with 
    \be \label{itr} I_{tr} = (D-2)\nabla^2\xi + \frac{\Box}{3}\tilde{\xi} - \Gamma + \dot{\tilde{\psi}}. \ee

    \no So we have the following set of spin-3 invariants under trDiff
    \be I_A^{trDiff} = \{I_{ijk}^{tt}, I_{ij}^{tt}, I_j^t, \tilde{I}_j^t, I, \tilde{I}, I_{tr}\}  . \label{trdinv} \ee

 \no   The reader can check that in (\ref{trdinv}) we have $N_I=D(D^2-1)/6 -1$ independent invariants, which agrees with (\ref{ni}). So we have a complete set of gauge invariants. In all gauge invariant theories that we have analysed so far, see \cite{Jaccard,da-hs,nstd}, the Lagrangian density is a quadratic  form on the gauge invariants,
  \be {\cal L} = I_{A}\, {\cal {M}}_{AB}\, I_B  . \label{matrix} \ee
 
\no where ${\cal{M}}_{AB}$ is a matrix differential operator. Indeed, up to total derivatives, we can rewrite the spin-3 Fronsdal Lagrangian in a explicitly gauge invariant way:
    \bea \nn \pazocal{L}_F &=& I_{ijk}^{tt}\frac{\Box}{2}I_{ijk}^{tt} - I_{jk}^{tt}\frac{3\nabla^2}{2}I_{jk}^{tt} + 3I_j^t\nabla^2\tilde{I_j}^t + \frac{3-3D}{2D}\tilde{I_j}^t\Box \tilde{I_j}^t \\ &+& \frac{1}{4}I_1(3\nabla^2+\Box)I_1 + \frac{3}{2}I_1\nabla^2\partial_0 I_{Tr}+\frac{3\nabla^2}{2}(D\!-\!2)I_1(\nabla^2+\Box)\tilde{I}\nn \\ &+& \frac{9}{2}(D\!-\!2)\tilde{I}\nabla^4 \partial_0 I_{Tr} + \frac{3\nabla^4}{4}(D\!-\!2)^2\tilde{I}\left(3\Box-\frac{D}{D-2}\nabla^2 \right)\tilde{I} +  \frac{9}{4}I_{Tr}\nabla^4 I_{Tr} , \label{f3i}\eea

    \no where, in order to simplify the $SO(D-1)$ scalar sector of ${\cal L}_F$, we have made the redefinition 
    \be I_1 \equiv I - 3\partial_0 I_{Tr} = \rho -3(D-2)\nabla^2\dot{\xi}-\nabla^2\dot{\tilde{\xi}} \label{i1} .\ee

    \no The independence of $(I_{ijk}^{tt},\tilde{I_j}^t )$ is clear from their association with $(\xi_{ijk}^{tt},\tilde{\xi}_j^t)$, respectively, see (\ref{invss3}) and (\ref{ijota}).   We can also associate $(I_{ij}^{tt},I_j^t)$ with redefinitions of $(\psi_{ij}^{tt},\gamma_j^t)$, respectively. In the $SO(D-1)$ scalar sector, one can associate the invariants $I_1$, $\tilde{I}$ and $I_{Tr}$ with $\rho$, $\psi$ and $\xi$, respectively.
    
    The physical content of the $SO(D-1)$ scalar sector can be checked by the determinant of the operators matrix
   \be \mathbb{M}_{\text{scalar}}=
\begin{bmatrix}

\frac{1}{4}\left(3\nabla^2 + \Box\right) & 
\frac{3\nabla^2(D-2)}{4}\left(\nabla^2 + \Box\right) & 
\frac{3}{4}\nabla^2 \partial_0 \\[2ex]

\frac{3\nabla^2(D-2)}{4}\left(\nabla^2 + \Box\right) & 
\frac{3\nabla^4(D-2)^2}{4}\left(3\Box - \dfrac{D}{D-2}\nabla^2\right) & 
\frac{9(D-2)}{4}\nabla^4 \partial_0 \\[2ex]

-\frac{3}{4}\nabla^2 \partial_0 & 
-\frac{9(D-2)}{4}\nabla^4 \partial_0 & 
\frac{9}{4}\nabla^4\\
\end{bmatrix} ,\ee
whose determinant is $\det(\mathbb{M}_{\text{scalar}})=-\frac{27}{16}(D-2)\nabla^{12}$, i.e., there is no propagating degrees of freedom in the scalar sector.

    Since $I_A^{trDiff}$ are all independent quantities, their equations of motion obtained from (\ref{f3i}) lead to the vanishing of all invariants: $I_A^{trDiff}=0$, except $I_{ijk}^{tt}$,
    which satisfies $\Box I_{ijk}^{tt} =0$ and represent, in $D=4$, the two $\pm 3$ propagating helicities. So we have a gauge invariant identification of the particle content of the spin-3 Fronsdal model as a preparation for the next section.
   
\section{Particle content via a shortcut}

In this section we show how, in some free gauge theories, the gauge invariant approach can save several steps in the identification of the particle content of the theory, turning it almost immediate. This will be very convenient in deriving the PBF model in section 4. Once again we take spin-2 and  spin-3 examples.

\subsection{Spin-2 TDiff model}

The minimal symmetry, see \cite{Ng}, for describing massless spin-2 particles via a rank-2 tensor is given by transverse diffeomorphisms (TDiff henceforth),
\be 
\delta h_{\mu\nu} = \del_\mu \eps^{T}_{\nu} +  \del_\nu \eps^{T}_{\mu}\, , \label{tdiffs} \ee

\no where $ \p^\mu \eps^T_\mu=0$. In searching for the TDiff invariants we recall that all Diff invariants are also TDiff invariants. However, the transverse condition reduces the number of gauge parameters by one unit which, according to (\ref{ni}), leads to one extra gauge invariant.  It is given by the trace since, $\delta \, h = 2 \, \p^{\mu}\eps_{\mu}^T =0$. The trace can not be a function of the $I_{A}^{Diff}$, since it is not Diff invariant. Therefore, we can define the  full set of TDiff invariants by $I_{\tilde{A}}^{TDiff}= \{I_{A}^{Diff},h\}$ where $I_{A}^{Diff}$ are given in (\ref{diffinv}) and (\ref{diffinvs}). On the other hand, the most general local field theory, second order in derivatives and TDiff invariant , see \cite{blas},  is given by
\be \mathcal L_{TD}(a,b) = \frac{1}{2} h^{\alpha\beta}\, \Box \,  h_{\alpha\beta}+(\p^\mu h_{\mu\beta})^2+ a\, h\, \p^{\mu}\p^{\nu} h_{\mu\nu}-\frac{b}{2} h\,  \Box h \, ,
\label{sab}
\ee

\no where $(a,b)$ are  arbitrary real constants. In the special case $(a,b)=(1,1)$ the TDiff symmetry is enhanced to Diff, it corresponds to the LEH theory with only healthy massless spin-2 particles in the spectrum. The particle content of (\ref{sab}) for arbitrary values of $(a,b)$ has been investigated in \cite{blas} via Lagrangian constraints and gauge invariants, in \cite{rrm} via the usual Dirac algorithm for constrained systems and in \cite{rr} via the analytic structure of the propagator. After a lengthy work, one concludes that the spectrum contains a spin-2 particle (always physical) and a spin-0 particle. The latter is physical (non physical) if $f_D(a,b)>0$ ($f_D(a,b)<0$). At $f_D(a,b)=0$ we only have the spin-2 particle. The quantity $f_D(a,b)$ is given by:
\be f_D(a,b) \equiv (a^2-b)(D-2) + (a-1)^2 . \label{fd} \ee

\no Remarkably, we can arrive at the same conclusions after just one step. Namely, there is a unique \cite{nstd} field redefinition which splits ${\cal L}_{TD}(a,b)$ into the sum of a LEH theory and a trace dependent term. Namely, after what we call an ``r-shift'', $h_{\mu\nu} \to h_{\mu\nu} + r\, \eta_{\mu\nu} h$, with $r=(1-a)/(a D-2)$, assuming $a\ne 2/D$, the TDiff theory (\ref{sab}) can be written, in terms of the new field, as 
\be {\cal L}_{TD}(a,b) = {\cal L}_{EH}(h_{\mu\nu}) + \frac{(D-2)\, f_D(a,b)}{2(a\, D-2)^2} \, h\, \Box \, h = {\cal L}_{EH}(I_A^{Diff})   + \frac{(D-2)\, f_D(a,b)}{2(a\, D-2)^2} \, h\, \Box \, h \, , \label{ehh} \ee

\no where ${\cal L}_{EH}(I_A^{Diff})={\cal L}_{EH}(I^{tt}_{ij},I_{i}^t,I_{00},I)$ is given in (\ref{lehinv}). Although the field $h_{\mu\nu}$  in ${\cal L}_{EH}(h_{\mu\nu})$ is not traceless, the content of ${\cal L}_{TD}(a,b)$ can be read off from the two terms in (\ref{ehh}) separately. Consequently, in full agreement with  \cite{blas, rrm, rr}, we have the physical massless spin-2 particle of the LEH theory plus a scalar particle represented by the trace $h$ which may be physical ($f_D>0$) or not ($f_D<0$). The key point is the independence of the invariants $I_A^{TDiff} = \{I^{tt}_{ij},I_{i}^t,I_{00},I,h\}$ as shown in detail in \cite{nstd}. For the singular case $a=2/D\equiv a_D$ there is no need of an r-shift in the fundamental field $h_{\mu\nu}$, as shown in \cite{nstd}, it is easy to split ${\cal L}_{TD}(a_D,b)$ into a WTDiff (Weyl plus TDiff) Lagrangian ${\cal L}_{TD}(a_D,b_D)$, with $b_D=(D+2)/D^2$ and a term proportional to $f_D(a_D,b)\, h \, \Box \, h$. The Lagrangian ${\cal L}_{TD}(a_D,b_D)$ is the linearized version of unimodular gravity with only massless spin-2 particle in the spectrum \cite{blas,nstd}, once again the sign of the kinetic term of the scalar field ($h$)  is determined by the sign of $f_D(a,b)$.

\subsection{Spin-3 TDiff model}

We have recently suggested \cite{3tdiff} a spin-3 analogue of (\ref{sab}), namely, 
\be \label{sabc} \pazocal{L}^{(3)} (a,b,c) =\! -\frac{1}{2}(\partial_\rho h_{\mu\nu\alpha})^2\! +\! \frac{3}{2}(\partial^\mu h_{\mu\nu\alpha})^2 + 3 \, a\ \partial^\mu \partial^\nu h_{\mu\nu\alpha} h^\alpha + \frac{3\,b}{2}(\partial_\rho h_\nu)^2 \!+\! \frac{3\,c}{4}(\partial^\mu h_\mu)^2 . \ee

\no For arbitrary values of the constants $(a,b,c)$ it is necessary that the gauge parameters  be both traceless and transverse (trTDiff) in order to be a symmetry, 
    \be \label{deltattd} \delta h_{\mu\nu\alpha} = \partial_{(\mu}\overline{\Lambda}_{\nu\alpha)}^T. \ee

    \no The invariants under traceless diffs are given in (\ref{trdinv}). The transverse condition $\p^{\nu}\overline{\Lambda}_{\nu\alpha}^T=0$ leads, according to (\ref{ni}), to $D$ extra invariants which can be identified with the trace $h_{\mu}$ as in the spin-2 case, since 
    $\delta h_{\mu}=\p_{\mu}(\eta\cdot\overline{\Lambda}^T) + 2 \, \p^{\nu}\overline{\Lambda}_{\nu\mu}^T=0$. The trace $h_{\mu}$ can not be a function of the traceless diff invariants since it is not symmetric under such transformations. Therefore, the trTDiff invariants can be taken as $I_{\tilde{A}}^{trTDiff}= \{I_{A}^{trDiff}, h_{\mu}\}$. Similarly to the spin-2 TDiff model, the particle content of (\ref{sabc}) can be read off immediately by splitting it into a Fronsdal model plus a trace dependent term via a local field redefinition. Assuming that $a\ne 2/(D+2)$, the required redefinition is unique again:
\be \label{redef2} h_{\mu\nu\alpha} \rightarrow h_{\mu\nu\alpha} + \frac{1-a}{a(D+2) - 2}\eta_{(\mu\nu}h_{\alpha)},\ee
\no after which we have
 \be \label{fv}\pazocal{L}^{(3)} (a,b,c) = \pazocal{L}_F(I_{ij}^{tt},I_{j}^t,\tilde{I}_j^t,I,\tilde{I},I_{Tr}) + \pazocal{L}_V (h_\mu), \ee

 \no where $\pazocal{L}_F(I_{ij}^{tt},I_{j}^t,\tilde{I}_j^t,I,\tilde{I},I_{Tr}) $ is the usual Fronsdal Lagrangian given in (\ref{f3i}) and the vector Lagrangian ${\cal{L}}_V$ is given only in terms of the trace,
   \be \label{lv}  {\cal{L}}_V = \frac{3D}{2[a(D+2)-2]^2}\left\lbrack f_b^{(3)}\, h_\nu \,\Box \,h^\nu + \frac{f_c^{(3)}}{2}(\partial^\mu h_\mu)^2 \right\rbrack,\ee

 \no where    
    \be  f_b^{(3)} = D\, (a^2-b) + (a-1)^2 \quad ; \quad 
   f_c^{(3)}  = D\, (c-a^2) + 2\, (a-1)^2 .  \label{fbc}
    \ee

\no Since the trTDiff invariant $h_{\mu}$ can be treated as an independent field coordinate, the particle content of (\ref{fv}) is the content of the usual Fronsdal theory (a physical massless spin-3 particle) plus the content of (\ref{lv}) which is in general a ghostly theory. It is known in massless HS models that the trTDiff symmetry (\ref{deltattd}) is not enough to get rid of non physical modes and must be enlarged.

If $f^{(3)}_c = 2\, f_b^{(3)}$, which fine tunes the constant $c$ by
\be c= 3\, a^2 - 2\, b  \equiv c_s (a,b) , \label{cs}\ee

   \no the vector Lagrangian becomes a Maxwell theory
    \be   {\cal{L}}_V = -\frac{3D \, f_b^{(3)}\,}{4[a(D+2)-2]^2} \, F_{\mu\nu}^2(h_{\mu}) . \label{max} \ee

  \no So, besides the physical massless spin-3 particle, we have a physical (non-physical) massless spin-1 particle if $f_b^{(3)}>0$ ($f_b^{(3)}<0$). Exactly the same spectrum at $c=c_S$ has been identified in \cite{3tdiff} in a much more laborious way from the analytic structure of the propagator from $\pazocal{L}^{(3)}(a,b,c_S) $. At $c=c_S$ and $f_b^{(3)}\ne 0 $, the traceless condition on the gauge parameter is lifted in general and we have a generalized TDiff symmetry, see \cite{3tdiff} for more details. In particular, it includes the Maxwell-like models of \cite{cf} where $a=b=c=0=c_S(0,0)$
  
  If, besides $c=c_S$, we also set $ b= b_S (a) \equiv  a^2 + (a-1)^2/D $ such that $f_b^{(3)}=0$, 
 we have the Fronsdal family ${\cal L}(a,b_S,c_S)$ which is continuously connected with the usual Fronsdal model ${\cal L}(1,1,1)$ via trivial r-shifts. The transverse condition on the gauge parameter is lifted and we have trDiff invariance and r-shifted versions thereof. 
 
 There is still another possibility of getting rid of the ghosts which gives rise to a new irreducible description of massless spin-3 particles, it will be worked out in the next subsection. Lastly, the case $a=2/(D+2)$ will be addressed elsewhere, see conclusion. It is similar to the $a=2/D$ case of spin-2 TDiff model. We can split it into a spin-3 WtrTDiff model \cite{sv} plus a vector Lagrangian without need of an r-shift.

\subsection{The spin-3 PBF model}

It is a simple exercise to show that a Lagrangian of the kind $(\p^{\mu}A_{\mu})^2 $ has no particle content. On the other hand, from (\ref{sabc}), (\ref{fv}) and (\ref{lv}) we see that if we fix $b=b_S(a)$ such that $f_b^{(3)}=0$, we have a Fronsdal model plus a term proportional to $f_c^{(3)} \, (\p^{\mu}h_{\mu})^2$. If $c\ne c_S$, which means  $f_c^{(3)}\,\ne 0$,  we stay away from the ``Fronsdal family'' while still describing only massless spin-3 particles without ghosts. So we are led to define a new massless spin-3 model:
\be {\cal L}^{(3)}_{PBF} (a,c) \equiv {\cal L}^{(3)} (a,b_S(a),c) \quad ; \quad c\ne c_S(a) . \label{3lpbf} \ee

For sake of simplicity, we  chose henceforth $a=1$, which implies
$b=b_S(1)=1$ while we leave $c$ arbitrary except for $c\ne 1$. So we have the one parameter set of new models:
\bea {\cal L}^{(3)}_{PBF} \! &=& \!\!\! -\frac{1}{2}(\partial_\rho h_{\mu\nu\alpha})^2\! +\! \frac{3}{2}(\partial^\mu h_{\mu\nu\alpha})^2 + 3 \,\partial^\mu \partial^\nu h_{\mu\nu\alpha} h^\alpha + \frac{3}{2}(\partial_\rho h_\nu)^2 \!+\! \frac{3c}{4}(\partial^\mu h_\mu)^2  \\ \label{11c} &=& {\cal L}_F(I_{\tilde{A}}^{trTDiff}) + \frac{3(c-1)}{4}(\partial^\mu h_\mu)^2, \, \label{11cb}\eea

\no which is invariant under doubly transverse traceless diffs (trDiff$^{\ddagger}$ henceforth),
    \be \label{deltanf} \delta h_{\mu\nu\alpha} = \partial_{(\mu}\bar{\Lambda}^\ddag_{\nu\alpha)} \quad ; \quad \eta^{\mu\nu}\bar{\Lambda}^\ddag_{\mu\nu} = 0 =\partial^\mu \partial^\nu \bar{\Lambda}^\ddag_{\mu\nu} = 0.\ee
    
\no The extra constraint  $\partial^\mu \partial^\nu \bar{\Lambda}^\ddag_{\mu\nu} = 0$ eliminates one gauge parameter and consequently, according to (\ref{ni}), we have one extra gauge invariant besides the traceless diff invariants $I_A^{trDiff}$ of the usual Fronsdal model. The extra invariant may be chosen as $\p^{\mu}h_{\mu}\equiv I_h$. Thus, for the PBF model, we can work with the set $I_{\tilde{\tilde{A}}}^{trDiff^{\ddagger}}=\{I_{\tilde{A}}^{trDiff}, I_h\}$. Next we show  why they can be considered as independent field coordinates. First we notice that
     $I_h = \dot{\rho} - \nabla^2 \Gamma - (D-2)\nabla^2\dot{\psi} - \nabla^2\dot{\tilde{\psi}} + (D-2)\nabla^4\xi + \nabla^4 \tilde{\xi}$. After the redefinitions $\bar{\Gamma} \equiv \Gamma - \dot{\tilde{\psi}} + \frac{1}{3}\ddot{\tilde{\xi}}$ and $\bar{\tilde{\psi}}\equiv \tilde{\psi} - \frac{\nabla^2}{3}\dot{\tilde{\xi}}$, we can write down the matrix  taking us from $\{\rho,\Gamma,\psi,\tilde{\psi},\xi,\tilde{\xi}\}$ to $\{ I, \bar{\Gamma},\tilde{I},\bar{\tilde{\psi}},I_{tr}, I_h\}$ as
     

    \be \mathbb{M}_{PBF} = \begin{bmatrix}
1 & -3\partial_0 & 0 & 3(\partial_0)^2 & 0 & -3(\partial_0)^3\\
0 & 1 & 0 & -\partial_0 & 0 & \frac{(\partial_0)^2}{3} \\
0 & 0 & 1 & 0 & -\p_0 & 0\\
0 & 0 & 0 & 1 & 0 & \frac{-(\partial_0)}{3} \\
0 & \nabla^{-2} & 0 & \nabla^{-2}\partial_0 & D-2 & \frac{\Box}{3\nabla^2}\\
\nabla^{-4}\partial_0 & -\nabla^{-2} & -(D-2)\nabla^{-2}\partial_0 & -\nabla^{-2}\partial_0 & D-2 & 1
\end{bmatrix}.\ee
    
     \no Remarkably, its determinant has no time derivative: $\, \det (\pazocal{M}_{PBF}) = 2(D-2)/3$. So we can write ${\cal L}_{PBF} = {\cal L}_F (I_A^{trTDiff}) + 3(c-1) I_h^2/4$. Consequently, on shell, all invariants $\{I_A^{trTDiff}, I_h\}$ vanish except the spin-$ 3 $ helicities  $\Box I_{ijk}^{tt}=0$.

\section{Integer spin-\texorpdfstring{$s$}{TEXT} PBF model}

\subsection{Definition}

The spin-$s$ analogue of (\ref{sab}) with trTDiff invariance is given by
\be \pazocal{L}^{(s)}(a,b,c) =  -\frac{1}{2} (\partial h)^2 + \frac{s}{2}(\partial\! \cdot\! h)^2 + \frac{a\,s(s-1)}{2}\ \partial\! \cdot\! \partial\! \cdot\! h\ h' + \frac{b\,s(s-1)}{4}\ (\partial h')^2  + \frac{c\,s!}{8(s-3)!} \ (\partial \cdot h')^2  \label{trtdiffgeral}\ee
for any $a,b,c \in \mathbb{R}$, with $h$ being a rank-s symmetric double-traceless tensor ($h^{''}=\eta\cdot\eta\cdot h=0$). As in the spin-3 case we implement an invertible r-shift field redefinition in order to split (\ref{trtdiffgeral}) into a usual spin-$s$ Fronsdal model plus extra terms that depend only upon the trace $h^{'}$. After\footnote{The double traceless property is preserved by the r-shift.} carrying out in (\ref{trtdiffgeral}):
    \be \label{transfqualquer} h_{\mu_1 ... \mu_s} \mapsto h_{\mu_1 ... \mu_s} + r\, \eta_{(\mu_1 \mu_2} h^{\alpha}_{\phantom \alpha \alpha \mu_3 ... \mu_{s-2})} , \ee
 \no the form of the action is preserved with the map $(a,b,c)\to \left(A(r),B(r),C(r)\right)$, where
 \bea \nn A(r) &=& a +r [a(D+2\, s-4)-2]; \\
        B(r) \nn &=& b[1+r(-4+D+2s)]^2 -r\{ 2+r(-6+D+2s)+2a[1+r(-4+D+2s)]\}; \\
        \label{alfabetogamo}C(r) &=& c[1+r(D-4+2s)]^2 + 2r\{2+rD+2rs-4a[1+r(D-4+2s)]\}. \eea

\no Assuming that\footnote{We comment later on that exception.} $a\ne a^* \equiv 2/(D+2\, s-4)$, there is a unique r-shift such that $A=1$. Namely, 
\be r = r(a,s)= \frac{1-a}{a(D+2\, s -4)-2} . \label{r} \ee

\no Notice that $r(a,s)$ is such that the r-shift is always invertible independently of $(a,s)$ as far as $a\ne a^*$. After the r-shift (\ref{r}) the  Lagrangian can be written as the usual  Fronsdal model  plus two trace dependent terms. Explicitly, 
\be \pazocal{L}^{(s)}(a,b,c) = {\cal L}_F^{(s)}(h) + \frac{s(s-1)(D+2s-6)}{4[a(D+2\, s -4)-2]^2}\left\lbrack f_b^{(s)} \, h^{'}\, \Box \, h^{'} + \frac{f_c^{(s)}(s-2)}2 \left( \p\cdot h^{'}\right)^2 \right\rbrack ,\label{BC} \ee

\no where ${\cal L}_F^{(s)}$
is the usual spin-s Fronsdal Lagrangian corresponding to $\pazocal {L}^{(s)}(1,1,1)$ . We have introduced two constants generalizing\footnote{Notice that $f_D(a,b)=f_b^{(2)}$, see (\ref{fd}).} $f_b^{(3)}$ and $f_c^{(3)}$,
\bea  
f_b^{(s)} &=& (D+2s-6)(a^2-b) +(a-1)^2 ;\nn \\  
f_c^{(s)} &=& (D+2s -6)(c-a^2)+2\, (a-1)^2  .
\eea 

 The corresponding action $S(a,b,c)$ is invariant, in general, under transverse and traceless Diffs (trTDiff). At the ``Fronsdal point'' $f_b^{(s)}=f_c^{(s)}=0$ we have a one parameter Fronsdal family invariant under a larger symmetry, i.e., traceless Diff (trDiff). The Fronsdal Lagrangian ${\cal L}_F^{(s)}$ must be a function of spin-$s$ trDiff invariants ${\cal L}_F^{(s)}={\cal L}_F^{(s)}(I_A^{trDiff})$. The transverse constraint on the traceless gauge parameters $\p \cdot \overline{\Lambda}=0$ is a traceless symmetric rank-(s-2) tensor just like the trace $h^{'}$, which is the natural candidate for the extra invariants raised up by the transverse condition according to (\ref{ni}). So there must exist, for arbitrary integer  spin-$s$, a basis of trTDiff invariants given by $I_{\tilde{A}}^{trTDiff}= \{I_{A}^{trDiff},h^{'}\} $. Since $h^{'}$ is not invariant under trDiff it must be independent of $I_{A}^{trDiff}$ and the content of (\ref{BC}) can be read off from the Fronsdal model and the last two trace terms independently as in the spin-3 case. 
 
 In particular, since $h^{'}$ is traceless, at the point $f_c^{(s)}=2 f_b^{(s)}$, which requires $c=c_S=3\, a^2 - 2\, b$, the $h^{'}$ terms inside brackets in (\ref{BC})  combine precisely into the irreducible Skvortsov-Vasiliev model \cite{sv} for massless spin-$(s-2)$ particles. So due to the independence of the invariants, the model (\ref{BC}) describes a massless spin-$s$ and a massless spin-$(s-2)$ particle at $c=c_S$. Indeed,  we know that at least at the point $c=c_S=a=b=0$ the model (\ref{trtdiffgeral}) becomes the rank-s Maxwell-like model of \cite{cf} constrained by the double traceless condition. According to \cite{cf,cubic} it should be a partially reducible model describing massless spin-$s$ and spin-$(s-2)$ particles, in agreement with our interpretation in terms of independent gauge invariants. This the analogue of the ``Maxwell point'' of the spin-3 TDiff model of last section. The same spectrum should hold in general for  $\pazocal{L}^{(s)}(a,b,c_S)$ such that $f_b^{(s)}\ne 0$.
 
 As in the spin-3 case, besides the ``Fronsdal point'' and the ``TDiff point'' $c=c_S$, there is another point where we expect (\ref{trtdiffgeral}) to be ghost free, which is our main concern. 
 Namely, at the ``PBF point''  $f_b^{(s)}=0$ with $f_c^{(s)}\ne0$, which requires $c\ne c_S$ and
 \be b= b_S^{(s)} = a^2 + \frac{(a-1)^2}{(D+2\, s - 6)} . \label{bss} \ee

 \no The theory becomes invariant under traceless Diff with vanishing double divergence (trDiff$^{\ddagger}$), $\delta h = \p\, \overline{\Lambda}^{\ddagger}$.  Henceforth we define the PBF model for arbitrary spin-$s$ and arbitrary $D$-dimensional Minkowski space as 
 \be {\cal L}_{PBF}^{(s)} = {\pazocal L}^{(s)}(a,b_S^{(s)},c) \quad ; \quad  c\ne  3\, a^2 -2\, b_S, \label{pbfdef}\ee

 \no with the further restriction $a\ne 2/(D+2s-4)$.
 
 The constraint $\p \cdot\p \cdot \overline{\Lambda}=0$ is a rank-(s-3) traceless tensor just like $\p\cdot h^{'}\equiv I_{h^{'}}$, which is trDiff$^{\ddagger}$ invariant but not trDiff invariant. So $I_{h^{'}}$ must be independent of $I_{A}^{trDiff}$. Thus, there must exist a basis of trDiff$^{\ddagger}$ invariants given by $I_{\tilde{A}}^{trDiff^{\ddagger}}= \{I_{A}^{trDiff}, I_{h^{'}} \} $. Assuming all those invariants as independent field coordinates, as we have shown in the spin-3 case, we would have on shell: $I_{h^{'}}=0$  altogether with the  usual spin-$s$ Fronsdal equations. So, the spin-$s$ helicities  are the only propagating modes. For sake of comparison with the Fronsdal model, in the next subsections we chose to deal with the PBF model ${\pazocal L}^{(s)}(1,1,c)$ with arbitrary $c\ne 1$. In the next subsection we look at its gauge independent equations of motion and argue that we can go further and consistently fix the higher spin de Donder gauge on shell. We show that we have the correct counting of degrees of freedom for arbitrary spin-$s$ in arbitrary $D$ dimensions. In the  subsection 4.3 we indeed prove, in the light cone gauge, the equivalence of the particle content of the PBF and Fronsdal models for  arbitrary integer spin-$s$.

 \subsection{Equations of motion and degrees of freedom}

The   equations of motion corresponding to the PBF model ${\pazocal L}^{(s)}(1,1,c)$, see (\ref{trtdiffgeral}),  correspond to the vanishing of the rank-s double-traceless Euler tensor:
\be E = \Box h - \p \ \pdot h + \eta \ \pdot \pdot h + \p \p h' - \eta \Box
h' -\frac{c}{2}\eta \ \p \ \pdot h' - \Upsilon \ \eta \ \eta \ \pdot \pdot h' =0, \label{eoms} \ee

\no where 
\be \Upsilon = \frac{1-c}{D+2(s-4)} . \label{gamao} \ee

\no The equations (\ref{eoms}) are invariant under trDiff$^{\ddagger}$:
\be \delta h = \p \, \overline{\Lambda}^{\ddagger} \quad . \label{ddt} \ee

\no In agreement with the gauge symmetry (\ref{ddt}), the divergence of the equations of motion is a linear combination of $\eta$ and two derivatives $\p\, \p$,
\be  \pdot E = \eta \left[\p^3\!\cdot \! h -\left(1 + \frac{c}{2}\right) \Box \pdot h' - \left(\Upsilon + \frac{c}{2}\right) \p \ \pdot \pdot h' - \Upsilon \eta \ \p^3 \! \cdot \! h'\right] +(1-c) \partial \ \partial \ \pdot h' =0.\\ \label{pdotk} \ee

\no Assuming vanishing fields at infinity, equations of the type $\eta\, X + \partial\p\, Y = 0$ have only the trivial solution $X=0=Y$. So we have the following constraints
\bea \pdot h'&=&0 , \label{c1} \\
\pdot \pdot \pdot h &=& 0. \label{c2}\eea

\no With those constraints, the trace of the equations of motion imply  $(\Box h' - \pdot \pdot h)=0$. Back in (\ref{eoms}) we obtain the vanishing of the Fronsdal tensor as in the Fronsdal model, i.e.,
\be {\pazocal F} = \Box h - \partial \ \pdot h + \p \ \p \ h' =0. \label{ft} \ee


All the equations  (\ref{eoms})-(\ref{ft}) are invariant  
under the gauge transformations (\ref{ddt}). Now we fix the 
spin-$s$ de Donder gauge,
\be \overline{f} = \p \cdot h - \frac 12 \p\, h^{'} =0  . \label{dedonder} \ee

\no The number of gauge conditions may not exceed the number of independent gauge parameters. Because of the gauge invariant constraints (\ref{c1}) and (\ref{c2}), the traceless rank-(s-1) symmetric tensor (\ref{dedonder}) has vanishing double divergence on shell: $\pdot\pdot \overline{f}=0$. Thus, we are effectively fixing on shell the same number of gauge conditions as the number of independent gauge parameters $\lbdd$. Using the de Donder gauge in (\ref{ft}) we obtain
the massless equations $\Box h = 0$. 

All the equations so far, including the gauge conditions (\ref{dedonder}) are invariant under residual harmonic ( $\Box \overline{\Lambda}^{\ddagger}=0$) gauge transformations 
 which allows us to ``regauge''. Now we proceed with the counting of degrees of freedom (d.o.f.). The double-traceless field $h$ has 
\be N_h = \binom{D+s-1}{D-1} - \binom{D+s-5}{D-1} \ee
independent components. The number of independent gauge parameters $N_\Lambda$ is 
\be N_{\Lambda} = \binom{D+s-2}{D-1} - \binom{D+s-4}{D-1} -\left [ \binom{D+s-4}{D-1} - \binom{D+s-6}{D-1} \right]. \ee
The equations of motion lead to the gauge independent constraints (\ref{c1}) and (\ref{c2}) which are traceless rank-(s-3) symmetric tensors. The number of independent constraints is
\be N_{\text{constraints}} = 2\left[ \binom{D+s-4}{D-1} - \binom{D+s-6}{D-1} \right] \ee
The number of physical degrees of freedom (d.o.f.), considering the ``regauging'' is then
\bea N_{dof} &=& N_h - 2 N_\Lambda - N_{\text{constraints}}\nn\\
 &=& \binom{D+s-1}{D-1} - 2\left[ \binom{D+s-2}{D-1} - \binom{D+s-4}{D-1} \right] - \binom{D+s-5}{D-1}. \label{ndof}\eea
For arbitrary $D$ dimensions and arbitrary spin-$s$, (\ref{ndof}) coincides exactly with the number of d.o.f. of the Fronsdal model.
The inclusion of higher derivative Lagrangian constraints in the counting of degrees of freedom is typical of gauge parameters with derivative restrictions, see chapter 5 of \cite{book}. This is the case also of the spin-2 TDiff model (\ref{sab}) whose equations of motion lead to the two derivative scalar constraint $\p^{\alpha}\p^{\beta}h_{\alpha\beta}=0$. So, although we have (in $D=4$)  only 3 independent gauge parameters due to the transverse condition $\p^{\mu}\eps^T_{\mu}=0$, we can fix the four gauge conditions $f_{\mu}=\p^{\mu}h_{\mu\nu}=0$ since they are on shell transverse due to the two derivative scalar constraint. Taking into account the ``regauging'' we have $N_{dof}= 10-2\times 3 -1=3$ in agreement with one spin-0 and the $\pm 2$ helicities of the spin-2 particle\footnote{Remarkably, the Hamiltonian analysis of \cite{rrm} has shown that the spin-2 TDiff model has $2D-1$, instead of the expected $2(D-1)$, first class constraints and no second class constraints. The Hamiltonian constraints analysis of the PBF model for arbitrary integer spin-$s$ is in progress.}.

\subsection{The PBF model in the light cone gauge}

Here we analyse the equations of motion of the PBF model $\pazocal{L}^{(s)}(1,1,c)$ in the light cone gauge where no residual symmetry is left over. For sake of clearness we take the $s=3$ and $s=4$ cases before the general case of arbitrary integer $s$. In the light cone we have  $(\eta_{-+},\eta_{+-},\eta_{IJ})= (-1,-1,\delta_{IJ})$ and $\eta_{++}=\eta_{--}=\eta_{\pm J}=0$, with $I,J = 1,2,\cdots, D-2$. 

\subsubsection{The spin-3 case}

We start with the gauge transformations of the spin-3 case in the momentum space:
\be \delta h_{\mu\nu\alpha} = i p_{(\mu} \lbdd_{\nu\alpha)} .\ee
The parameters $\lbdd_{\nu\alpha}$ satisfy the restrictions:
\be p\! \cdot \! p \! \cdot \! \lbdd = 0 = \eta\cdot\lbdd. \label{conds}\ee
Assuming $p_+\neq0$, due to (\ref{conds}), the gauge parameters $\lbdd_{--}$ and $\lbdd_{+-}$ become functions of $(\lbdd_{++},\lbdd_{IJ})$. However, we can use the whole set of unconstrained parameters $(\lbdd_{++},\lbdd_{IJ},\lbdd_{+J},\lbdd_{-J})$ in order to impose the maximal number of  gauge conditions:
\be h_{+++}=0=h_{+IJ}=h_{++J}=h_{+-J}\, .  \label{gc3+} \ee

\no As mentioned before, the PBF equations of motion, even before  any gauge fixing, also lead to the Fronsdal equations:
\be \label{ftmomentum} E_{\mu\nu\rho}(p) \equiv p^2 h_{\mu\nu\alpha} - p_{(\mu}p^\rho h_{\nu\alpha)\rho} + p_{(\mu} p_\nu h_{\alpha)} = 0 .\ee

\no Henceforth we use the short hand notation: $(\mu,\nu,\rho)\equiv E_{\mu\nu\rho}(p)$ and assume that (\ref{gc3+}) is holding true. From $(+,+,+)=0$ one finds $h_{-++}=0$. The equations $( +,+,J)=0$ and $(+,I,J)=0$ lead to $h_{IIJ}=0$ and $h_{-IJ} = \frac{1}{p^+}p_K h_{IJK}$, respectively, they lead altogether to $h_{-JJ} =0$. The constraint $\partial\!\cdot\!h'=0$ implies $h_{+--}=0$. The field is thus on shell traceless ($h_\alpha= -2\, h_{+-\alpha} + h_{II\alpha} =0$) and satisfies  $h_{+\mu\nu}=0$. From $(+,\nu,\alpha)=0$ we have the transverse constraint $p^{\rho}h_{\rho\nu\alpha}=0$ which at $(\nu,\alpha)=(-,k)$ and $(\nu,\alpha)=(-,-)$ leads, using previous results, to 
\be h_{--K}=\frac{p_I\, p_J\, h_{IJK}}{p_+^2} \quad ; \quad h_{---}=\frac{p_I\, p_J\, p_K h_{IJK}}{p_+^3} . \label{hmenos} \ee

\no So we can consider $h_{IJK}$ as the only relevant field components. With $h_{\alpha}=0=p^{\rho}h_{\rho\mu\nu}$ back in (\ref{ftmomentum}) we have $p^2 h_{\mu\nu\alpha} = 0$. Effectively, 
we have the following physical equations:
\be p^2h_{IJK}=0, \quad h_{JJK}=0. \label{3eomlc}\ee

\no Thus, we have $(D+2)(D-2)(D-3)/6$ massless modes in $D$ dimensions. As expected we have $(0,2,7)$ modes in $D=(3,4,5)$ respectively since, besides the expected helicities $\pm 3$ in $D=4$, in $D=3$ the Fronsdal model is known to be trivial and in $D=5$ we should have the same number ($2\, s +1$) of massive modes in $D=4$.
The condition $\p^{\mu}\p^{\nu}\lbdd_{\mu\nu}=0$ is compensated by the constraint $\p^{\mu}h^{'}_{\mu}=0$.

\subsubsection{The  spin-4 case.}

\no First, we notice that the traceless condition $\eta^{\mu\nu}\lbdd_{\mu\nu\rho}=-2 \, \lbdd_{+-\rho}+\lbdd_{KK\rho}=0$
and the vanishing double divergence  $\p^{\mu}\p^{\nu}\lbdd_{\mu\nu\alpha}=0$ constrain the parameters $\lbdd_{+-\rho}$ and $\lbdd_{--\alpha}$ respectively as functions of other gauge parameters. So they can not be used to fix gauges. Furthermore, with $\rho=-$ and $\alpha=+$, they simultaneously constrain $\lbdd_{--+}$ which, see $\eta\cdot\lbdd_{-}=0$,  amounts to fix both $\lbdd_{--+}$ and $\lbdd_{KK-}$. This is important to account for the gauge freedom at our disposal. Now, using the unconstrained parameters $\{\lbdd_{+++}, \lbdd_{++j}, \lbdd_{+ij}, \lbdd_{ijk}\}$ we can fix the gauge
\be h_{++++} = 0=  h_{j+++}=h_{ij++}=h_{ijk+} \quad . \label{gc4}\ee

 \no On the other hand, the gauge invariant
 equations of motion in the momentum space are given by
\be p^2h_{\mu\nu\alpha\beta} - p_{(\mu} \ p^{\rho}  h_{\rho\nu\alpha\beta)} + p_{(\mu}\ p_{\nu} \ h'_{\alpha\beta )}\equiv (\mu,\nu,\alpha,\beta) =0.\label{eoms4} \ee
From $(++++)=0$ one obtains $h_{+++-}=0$. The equations $(+++I)=0$ lead to $h_{++-I}=0$, and  $(++IJ)=0$ gives 
\be h_{KKIJ}=0 . \label{trs4}\ee

\no From  $(-+++)=0$ and the constraint  $(p\cdot h^{'})_+ =0$, we have $h_{++--}=0=h_{+-KK}$. Now we make use of the only unconstrained gauge parameters left over. Namely, since $\delta h_{+-IJ}= i\, p_{(+}\lbdd_{IJ-)}$ and we already have $h_{+-KK}=0$, we only need the space traceless parameter $\lbdd_{\overline{I}\overline{J}-}= \lbdd_{IJ-} - \delta_{IJ}\lbdd_{KK-}/(D-2)$ in order to fix our last light cone gauge condition: $h_{+-\overline{I}\overline{J}}=0$. Consequently, we have $h_{+-IJ}=0$ henceforth. There is no remaining unconstrained parameter anymore, since $\lbdd_{KK-}$ is already constrained as we have argued before (\ref{gc4}).

Next, from $(++-I)=0$ we have $h_{-KKJ}=0$. So the constraint $(p \cdot h^{'})_I=0$, leads to $h_{+--I}=0$. The remaining constraint $(p \cdot h^{'})_-=0$ and previous results lead to a vanishing trace $h_{\mu\nu}=0$ on shell. From $(++--)=0$ we have $h_{+---}=0$ and therefore $h_{+\mu\nu\rho}=0$. Back in $(+\mu\nu\rho)=0$ we have the on shell transverse constraint $p^{\alpha}h_{\alpha\mu\nu\rho}=0$, from which we can eliminate all non vanishing components with at least one ``-'' index, i.e., 
\bea h_{-IJK}=\frac{p_L\, h_{IJKL}}{p_+} \quad &;& \quad h_{--IJ}=\frac{ p_K\,  p_L\, h_{IJKL}}{p_+^2} \nn\\h_{---I}=\frac{p_J\, p_K\, p_L\, h_{IJKL}}{p_+^3} \quad &;& h_{----}=\frac{p_I \, p_J\, p_K\, p_L\, h_{IJKL}}{p_+^3} \label{hmenoss4} \eea

\no From the transverse constraint back in $(IJKL)=0$ we obtain the massless condition for the propagating degrees of freedom,
\be p^2 h_{IJKL} = 0 . \label{s4we}\ee

\no Altogether with (\ref{trs4}) we have $(D+4)(D-3)(D-2)(D-1)/(4 \ !)$
massless propagating modes as in the Fronsdal spin-4 model.

\subsubsection{Arbitrary integer spin}

Now we introduce the notation $\left(+(l)-(m)J(n)\right)$  indicating that $J(n)=J_1J_2\cdots J_n$ and that we have $l$ indices $+$ and $m$ indices $-$. The traceless condition $(\eta\cdot\lbdd)_{\rho_1 \cdots \rho_{s-3}}=0$ constrains all gauge parameters with at least one ``+'' and one ``-'' index, namely,
$2\lbdd_{+-\rho_1 \cdots \rho_{s-3}}=\lbdd_{JJ\rho_1 \cdots \rho_{s-3}}$. The vanishing double divergence $(\p \cdot \p\cdot \lbdd)_{\alpha_1 \cdots \alpha_{s-3}}=0$ constrains $\lbdd_{--\alpha_1 \cdots \alpha_{s-3}}$. The overlapping cases where at least one of the $\rho_j$ is ``-'' and one of the $\alpha_j$ is ``+'' will restrict $ \lbdd_{JJ-\beta_2 \cdots \beta_{s-3}}$. So we can use the unconstrained gauge parameters $\lbdd_{+(k-1)j(s-k)}$ with $k=1, \cdots,s$ in order to fix the gauge
\be h_{+(k)J(s-k)}=0 \quad , \quad k=1, \cdots,s  \, . \label{gcs} \ee

\no The Fronsdal equations in momentum space read 
\be   p^2 \, h_{\mu_1\cdots \mu_s} + p_{(\mu_1}p_{\mu_2}\, h^{'}_{\mu_3\, \mu_4\cdots \mu_s)} - p_{(\mu_1} p^{\rho}h_{\rho\, \mu_2 \cdots \mu_s)} \equiv \left(\mu_1\, \mu_2 \cdots \mu_s\right) = 0.\ee

\no  From the equations of motion $\left(+(s-k)J(k)\right)=0$ we obtain $h_{-+(s-1-k)J(k)}=0$, where
$k=0,1, \cdots,s-3$. After a careful account of the combinatorial factors, from $\left(++J(s-2)\right)=0$ we have
\be h_{KKJ_1 \cdots J_{s-2}} = 0 . \label{trs} \ee

\no Now we can insert one ``-'' index. From $\left(-+(s-1-k)J(k)\right) = 0 $  we have $h_{--+(s-2-k)J(k)}=0$ where $k=0,1, \cdots,s-5$. From 
$\left(-+++J(s-4)\right) = 0 $ and the constraint $(p\cdot h^{'})_{+J(s-4)}=0$ we obtain $h_{--++J(s-4)}=0$ and $h_{-+KKJ(s-4)}=0$. Now we can use the unconstrained space traceless parameter 
$\lbdd_{-\overline{k_1}\overline{k_2}J(s-4)}$ and fix the gauge condition $h_{-+\overline{k_1}\overline{k_2}J(s-4)}=0$ such that we have $h_{-+J(s-2)}=0$. We proceed by adding ``-'' indices. For $N\ge 3$,   from $\left(-(N)+(s-k-N)J(k)\right) = 0 $  we can show that $h_{+(s-k-N-1)-(N+1)J(k)}=0$ with $k=0,1, \cdots, s-N-4$. For the case $k=s-N-3$ we need also the constraint $(p\cdot h^{'})_{+-(N-1)J(s-N-3)}=0$ which altogether with  $\left(-(N)+++J(s-N-3)\right) = 0 $ leads to $h_{-(N+1)++J(s-N-3)}=0$ and $h_{+-(N)KKJ(s-N-3)}=0$. So we end up with 
\be h_{++\mu (s-2)}=0 \quad \, \text{and} \, \quad h_{+KK\mu(s-3)}=0 , \label{hpps} \ee

\no From $\left(++-J(s-3)\right)=0$ we have $h_{-KKJ(s-3)}=0$, back in the constraint $(p\cdot h^{'})_{J(s-3)}=0$ we have $h_{+--J(s-3)}=0$. We keep on adding ``-'' indices and subtracting ``J'' space indices such that we deduce $h_{+-(k)J(s-k-1)}=0$ with $0\le k \le s-1$ and consequently,
\be h_{+\mu (s-1)}=0 . \label{hps} \ee

\no We have again a vanishing trace on shell $h^{'}=0$. From  $\left(+\mu_1\cdots\mu_{s-1}\right)=0$ we have the on shell transverse condition $p^{\rho}h_{\rho\mu_1 \cdots \mu_{s-1}}=0$ from which we can eliminate all non vanishing field components with at least one ``-'' index:
\be \label{arblightconeminus} h_{-(l)I(s-l)} = \frac{1}{(p_+)^l}p_{I_1}p_{I_2}\dots p_{I_l} h_{I_1 I_2 \dots I_l I(s-l)}, \ee
where $l \in \{1, 2, \dots, s \}$. Notice that the constraint $p \cdot p\cdot p\cdot h = 0$ is equivalent to (\ref{arblightconeminus}). So the only nontrivial field components are given by $h_{J(s)}$. The equations of motion $\left(I_1 \cdots I_s\right)=0$ become simply
\be p^2 h_{I(s)}=0. \ee

\no Since we still have the algebraic constraints (\ref{trs}) we end up with two helicities in $D=4$. The condition on the gauge parameter $\p \cdot\p\cdot\lbdd=0$ is compensated by the constraint $\p\cdot h^{'}=0$. Notice that they are both totally symmetric rank-(s-3) traceless tensors.

\section{Conclusion and perspectives}

 We have found here a partially broken Fronsdal (PBF) model for arbitrary integer spin,
it is defined in (\ref{pbfdef}), (\ref{bss}) and (\ref{trtdiffgeral}). The model is invariant under traceless higher spin diffeomorphisms with vanishing double divergence, see (\ref{ddt}). In subsection 3.3 we have verified the content of the spin-3 PBF model in a gauge invariant way while in subsection 4.3 we have  checked the spectrum of the arbitrary integer spin case with $s\ge 3$ in the light cone gauge. We conclude that the PBF model has the same propagating content of the Fronsdal model, i.e.,  only the $\pm s$ helicities in $D=4$. 

The reduced symmetry of the PBF model however, leads to less constraints on the source, for instance, in $D=4$, in the spin-3 case we have 8 constraints instead of the 9 ones of the Fronsdal theory, so more general couplings are allowed. Explicitly, the gauge invariance of the source term: $\delta \int T\, h = -s\, \int (\p\cdot T) \, \overline{\Lambda}^{\ddagger} = 0$ implies
\be \p \cdot T = \eta\, \Omega + \p\, \p \, J , \label{source} \ee
\no where $\Omega$ and $J$ are rank-(s-3) symmetric tensors constrained only by the fact that the double trace of the right hand side  of (\ref{source}) must vanish for $s \ge 5$. In the Fronsdal theory the non conservation law is more restrictive ($J=0$). 

As we have mentioned in the introduction, there is another formulation \cite{sv} of arbitrary integer spin-$s$ massless particles. It is invariant under transverse and traceless higher spin diffeomorphisms plus traceless Weyl-like transformations : $\delta h = \p \, \overline{\Lambda}^T + \eta\, \overline{\sigma} $. Comparing with the Fronsdal model where $\delta h = \p \, \overline{\Lambda}$ we see that the transverse condition is compensated by the Weyl symmetry such that the gauge transformations both in Fronsdal and \cite{sv} have the same number of gauge parameters and consequently the same number of constraints on the source. It is also possible to define a partially broken version of \cite{sv}  where the traceless Weyl parameter is further constrained to be transverse $\p\cdot \overline{\sigma}^T=0$. The model so defined also describes only spin-$s$ helicities as propagating modes. It is connected with the special point $a=2/(D+2s-4)$ that we have not considered in the present work and it will be addressed elsewhere. For the specific case of $s=3$ and $D=4$, the partially broken version of \cite{sv} has recently appeared in the literature, see model ${\mathfrak{E}}_6$ of \cite{bms3}. It turns out that both transformations $\delta h = \p \lbdd$ and $\delta h =  \p \, \overline{\Lambda}^T + \eta\, \overline{\sigma}^T$ have the same number of gauge parameters which is apparently the minimal number necessary for describing massless integer spin-$s$ particles.

Regarding a curved background, since $\nabla\cdot h^{'}$ is by itself invariant under $\delta h = \nabla \overline{\Lambda}^{\ddagger}$, with $g^{\mu_1\mu_2}\, \overline{\Lambda}^{\ddagger}_{\mu_1\mu_2\cdots \mu_{s-1}}=0=\nabla \cdot \nabla \cdot \overline{\Lambda}^{\ddagger} $, it follows that the same non minimal compensating terms of the Fronsdal action \cite{fronsdal2} on (A)dS backgrounds, quadratic on $h$ and $h^{'}$, will turn the model $\pazocal{L}^{(s)}(1,1,c)$ given in (\ref{transfqualquer}), with $\p_{\mu} \to \nabla_{\mu} $,
 invariant under  gauge transformations on (A)dS. 

Concerning the relationship of the PBF model with the tensionless limit of String theory, we recall that the works \cite{fs1,fs2,st,fscqg} have found an interesting connection between the Fronsdal theory where both the field (double traceless) and the parameter (traceless) are constrained and string theory in the tensionless limit  where neither the fields nor the gauge parameters are constrained. The connection is established via a reducibility condition on the so called triplet equations stemming from string theory. It might be interesting to introduce two rank-(s-3) fields in the PBF model in order to undo the constraints ($\eta\cdot\lbdd =0=\p\cdot\p\cdot\lbdd$) altogether with a rank-(s-4) Weyl symmetry to undo the double traceless condition along the lines of \cite{more} and search for a convenient reducibility constraint.

It is important to mention that the PBF model is a two parameters theory, $\pazocal{L}_{PBF}^{(s)}=\pazocal{L}^{(s)}[a,b_S^{(s)}(a,c),c]=\pazocal{L}_{PBF}^{(s)}(a,c)$. We can always redefine one of the parameters as we wish by means of local field redefinitions that we have named r-shifts, see (\ref{transfqualquer}) and (\ref{alfabetogamo}). We are currently investigating the role of the remaining parameter which can not be shifted, along the lines of \cite{3tdiff} where we have shown, inspired also by non-local field redefinitions, that the two parameters TDiff spin-3 model $S[a,b,c_S(a,b)]$, see (\ref{sabc}), is physically equivalent to its simplest version $S(0,0,0)$ which corresponds to the spin-3 case of the Maxwell-like model of \cite{cf}. Similarly, we believe that non-local field redefinitions may help to clarify the role of the left over constant parameter of the PBF model.


Last and not least, it is natural to search for the consequences of the reduced gauge symmetry regarding self-interacting vertices as well as a fermionic generalization of the PBF model to be compared with the known Fang-Fronsdal models \cite{fang}. This is in progress.

\section{Acknowledgements}

The work of D.D. is partially supported by CNPq (grant 306380/2017-0) while BSM was
supported by CAPES and is currently supported by FAPESP (grant 2025/04628-1). We thank Elias L. Mendonça and Raphael S. Bittencourt for several discussions. We thank Raphael S. Bittencourt also for bringing \cite{bms3} to our knowledge.


\begin{thebibliography}{}

\bibitem{dirac} Dirac P A M 1936 Proc. R. Soc. Lond. A 155 447.

\bibitem{fp} M. Fierz,  1939 Helv. Phys. Acta 12 3,
M. Fierz and W. Pauli, 1939 Proc. R. Soc. Lond. A 173 211.

    
\bibitem{snow}
	X.~Bekaert, N.~Boulanger, A.~Campoleoni, M.~Chiodaroli, D.~Francia, M.~Grigoriev, E.~Sezgin, and E.~Skvortsov,
	``Snowmass White Paper: Higher Spin Gravity and Higher Spin Symmetry,''
	[arXiv:2205.01567 [hep-th]].
	
\bibitem{rakibur_r} R.~Rahman and M.~Taronna,
		``From Higher Spins to Strings: A Primer,''
		[arXiv:1512.07932 [hep-th]].
		
\bibitem{sagnoti_r1}	 A. Sagnotti, “Notes on Strings and Higher Spins,” J. Phys., vol. A46, p. 214006, 2013, arXiv:1112.4285.
		
	
\bibitem{vasiliev_r} X. Bekaert, S. Cnockaert, C. Iazeolla and M. A. Vasiliev, “Nonlinear higher spin
theories in various dimensions,” arXiv:hep-th/0503128.

\bibitem{sagnoti_r}	 A. Sagnotti and M. Taronna, ``String Lessons for Higher-Spin Interactions'', Nucl.Phys.,
 vol. B842, pp. 299–361, 2011,  e-Print:1006.5242 [hep-th]. 
 
\bibitem{book}
	A.~Bengtsson,
	``Higher Spin Field Theory, Vol 1+2,''
	De Gruyter, 2023.

\bibitem{fronsdal}
	C.~Fronsdal,
	``Massless Fields with Integer Spin,''
	Phys. Rev. D \textbf{18}, 3624 (1978),
	doi:10.1103/PhysRevD.18.3624.

    \bibitem{fs1}  D. Francia and A. Sagnotti, Free geometric equations for higher spins, Phys. Lett. B 543 (2002) 303 [hep-th/0207002].
	
\bibitem{sv}
	E.~Skvortsov and M.~Vasiliev,
	``Transverse Invariant Higher Spin Fields,''
	Phys. Lett. \textbf{B664}, 301 (2008),
	[arXiv:0701278 [hep-th]].
	
\bibitem{porrati}
	M.~Porrati,
	``Universal Limits on Massless High-Spin Particles,''
	Phys. Rev. D \textbf{78}, 065016 (2008),
	doi:10.1103/PhysRevD.78.065016,
	[arXiv:0804.4672 [hep-th]].
	
\bibitem{bbs}
	X.~Bekaert, N.~Boulanger, and P.~Sundell,
	``How Higher-Spin Gravity Surpasses the Spin-Two Barrier: No-Go Theorems Versus Yes-Go Examples,''
	Rev. Mod. Phys. \textbf{84}, 987–1009 (2012).
	
 \bibitem{Jaccard}
    M. ~Jaccard, M. ~Maggiore, and E. ~Mitsou,
    ``Bardeen variables and hidden gauge symmetries in linearized massive gravity'' Phys. Rev. D.\textbf{87}, 044017 (2013).
    
\bibitem{da-hs}D. Dalmazi and A.L.R. dos Santos, ``Higher spin analogs of linearized topologically massive gravity and linearized new massive gravity'',
Phys.Rev.D \textbf{104} 8, 085023 (2021), arXiv:2107.08879 [hep-th].

\bibitem{nstd}D. Dalmazi and L.G.M. Ramos, ``Transverse diffeomorphism invariant spin-2 theories via gauge invariants'', e-Print: 2506.24047 [hep-th] (2025). 

 \bibitem{Ng}
	J.~J.~van der Bij, H.~van Dam, and Y.~J.~Ng,
	``The Exchange of Massless Spin-Two Particles,''
	Physica \textbf{116A}, 307-320 (1982).
	
\bibitem{blas}
	E.~Alvarez, D.~Blas, J.~Garriga, and E.~Verdaguer,
	``Transverse Fierz–Pauli Symmetry,''
	Nucl. Phys. \textbf{B756}, 148 (2006);
	D.~Blas,
	``Aspects of Infrared Modifications of Gravity,''
	PhD Thesis, University of Barcelona, 
	[arXiv:0809.3744].
	
\bibitem{rrm} R. R. L. Santos. ``Transverse diffeomorphism and spin-2 particles'' (master dissertation), July 2020. 
https://repositorio.unesp.br/items/301c5fa8-da68-49f8-bc52-52122ad9ca04.
 
\bibitem{rr}
	D.~Dalmazi and R.~R.~L.~d.~Santos,
	``Eur. Phys. J. C,'' \textbf{81} (2021) no.6, 547,
	doi:10.1140/epjc/s10052-021-09297-0,
	[arXiv:2010.12051 [hep-th]].
	
 \bibitem{3tdiff} R. S. Bittencourt, D. Dalmazi, B.dos S. Martins, E.L. Mendonça,``Equivalence of spin-2 and spin-3 models invariant under transverse diffeomorphisms and the tensionless limit of string theory'', Phys.Rev.D 112 (2025) 2, 025008, e-Print:2501.04214.
 
\bibitem{cf}
	A.~Campoleoni and D.~Francia,
	``JHEP,'' 03 (2013) 168,
	[arXiv:1206.5877 [hep-th]].

     \bibitem{cubic}
	D.~Francia, G.~L.~Monaco, and K.~Mkrtchyan,
	``Cubic Interactions of Maxwell-Like Higher Spins,''
	JHEP \textbf{04}, 068 (2017),
	[arXiv:1611.00292 [hep-th]].

    \bibitem{bms3} Will Barker, Carlo Marzo and Alessandro Santoni, ``Infrared foundations for quantum geometry I: Catalogue of totally symmetric rank-three field theories'',     e-Print: 2506.21662.

\bibitem{fronsdal2} C. Fronsdal, Phys. Rev. D 20 (1979) 848.

\bibitem{fs2}	D. Francia and A. Sagnotti, Minimal local Lagrangians for higher-spin geometry,
Phys. Lett. B 624 (2005) 93 [hep-th/0507144].

	\bibitem{st}
	A.~Sagnotti and M.~Tsulaia,
	``On Higher Spins and the Tensionless Limit of String Theory,''
	Nucl. Phys. \textbf{B682}, 83 (2004),
	[arXiv:hep-th/0311257].
	
\bibitem{fscqg}
	D.~Francia and A.~Sagnotti,
	``On the Geometry of Higher Spin Gauge Fields,''
	Class. Quant. Grav. \textbf{20}, S473 (2003),
	[arXiv:0212185 [hep-th]].

   

 \bibitem{more}  R. Schimidt Bittencourt, D. Dalmazi and E.L. Mendonça   Phys.Rev.D 111 (2025) 2, 025022, e-Print: 2501.01596.
 
 \bibitem{fang} J. Fang, C. Fronsdal, Phys. Rev. D 18 (1978) 3630.
\end{thebibliography}
\end{document}